\documentclass[aps,prl,twocolumn,superscriptaddress,showpacs,aps]{revtex4}

\usepackage{amsmath,epsfig,amsfonts,graphicx}
\usepackage{subfigure}
\usepackage{graphicx}
\usepackage[american]{babel}
\usepackage{amssymb}
\usepackage{amsfonts}
\usepackage{color}
\usepackage[normalem]{ulem}

\begin{document}

\title{Rogue waves in nonlocal media}

\author{Theodoros P. Horikis}
\affiliation{Department of Mathematics, University of Ioannina, Ioannina 45110, Greece}

\author{Mark J. Ablowitz}
\affiliation{Department of Applied Mathematics, University of Colorado, 526 UCB, Boulder, CO
80309-0526, USA}

\begin{abstract}
The generation of rogue waves is investigated via a nonlocal nonlinear Schr\"odinger (NLS)
equation. In this system, modulation instability is suppressed and is usually expected that rogue
wave formation would also be limited. On the contrary, a parameter regime is identified where the
instability is suppressed but nevertheless the number and amplitude of the rogue events increase,
as compared to the standard NLS (which is a limit of the nonlocal system). Furthermore, the
nature of these waves is investigated; while no analytical solutions are known to model these
events, numerically it is shown that they differ significantly from either the rational
(Peregrine) or soliton solution of the limiting NLS equation. As such, these findings may also
help in rogue wave realization experimentally in these media.
\end{abstract}


\maketitle

Many destructive phenomena in the oceans have been identified as the result of rogue waves, extreme
waves that grow abnormally during a given sea state. This has motivated  wide ranging research on
rogue and extreme phenomena that spans across sciences
\cite{solli,chabchoub,shats,pisarchik,finance,bec1,zaviyalov}. Key to understanding these phenomena
is their common description through the well-known nonlinear Schr\"odinger equation (NLS) with
cubic/Kerr nonlinearity \cite{waves}.

The NLS system provides a unique balance between the critical effects that govern propagation in
dispersive media, namely dispersion/diffraction and nonlinearity. This balance leads to the
formation of solitons, which are characterized by their stability and robustness as they maintain
their shape and velocity even when they interact. Rogue waves, on the other hand, are observed to
appear from nowhere and are short-lived. They are often modeled by the so-called Peregrine soliton
\cite{peregrine,breather}, which is a special type of solitary wave formed on top of a continuous
wave (cw) background and in contrast to other soliton solutions of the NLS equation it is written
in terms of rational functions with the property of being localized in both time and space. These
properties make these solutions useful to describe such events \cite{akhmediev2,baronio2}.

The specific conditions that cause their formation is still a subject of enormous interest; it is
generally recognized that modulation instability (MI) is among the most important mechanisms which
lead to rogue wave excitation
\cite{zakharov,zakharov2,akhmediev,onorato_report,baronio3,erkintalo}. MI is the nonlinear
mechanism of the self-wave interactions, called the Benjamin-Feir instability \cite{bf} in water
wave physics. In nonlinear optics, it is considered as a basic process that classifies the
qualitative behavior of modulated waves \cite{mi}. Rogue waves, as a result of an MI process, can
be identified as high-contrast peaks of random intensity and are the result of the unstable growth
of weak wave modulations. Mathematically, MI is a fundamental property of many nonlinear dispersive
systems and is a well documented and understood phenomenon \cite{dudley}.

The NLS equation provides important information about rogue phenomena. However, for several
physically relevant contexts the standard focusing NLS equation turns out to be an oversimplified
description as it cannot model, for example, gain and loss which are inevitable in any physical
system \cite{agrawal}. Hence, in order to model important classes of physical systems in a relevant
way, it is necessary to go beyond the standard NLS description. There are, for example, important
systems that display nonlocal nonlinear mechanisms. Such media include, nematic liquid crystals
\cite{ass0,assanto_book}, thermal nonlinear optical media \cite{rot,kroli2} and plasmas
\cite{litvak,plasma}.

The effect of the nonlocality on the NLS equation is rather profound. The integrable nature of the
equation is generally lost and while soliton solutions may also be found they lack a free parameter
linking their amplitude to their velocity. In terms, of rational (rogue type) solutions none are
known, to our knowledge. In terms of the MI properties in the  model we investigate, the cw
solutions are always unstable with the nonlocality suppressing the instability (although it does
not eliminate the effect) \cite{kroli}. In fact, the effect is so strong that it has been proven
that the nonlocality eliminates collapse in all physical dimensions \cite{bang}. These observations
suggest that the nonlocality has a stabilizing effect. As such, it might be expected that rogue
wave phenomena should be more scarce and more prominent in the weakly nonlocal regime where the
system is closer to the standard NLS equation. We find, here, that the nonlocality does not always
suppress the number and size of the rogue events.

The normalized system that governs propagation in nonlocal media reads
\cite{kivshar_book,assanto_book}
\begin{subequations}
\begin{gather}
i\frac{{\partial u}}{{\partial z}} + d\frac{{{\partial ^2}u}}{{\partial {x^2}}} + 2g\theta u
= 0 \\
\nu \frac{\partial ^2\theta }{\partial x^2} - 2q\theta  =  - 2|u|^2
\end{gather}
\label{nls}
\end{subequations}
Depending on the physical situation the system and its coefficients correspond to different
physical quantities. For example, in the context of nematic liquid crystals, $u$ is the complex
valued, slowly varying envelope of the optical electric field and $\theta$ is the optically induced
deviation of the director angle. Diffraction is represented by $d$ and nonlinear coupling by $g$.
The effect of nonlocality $\nu$ measures the strength of the response of the nematic in space, with
a highly nonlocal response when $\nu$ is large. The parameter $q$ is related to the square of the
applied static field which pre-tilts the nematic dielectric \cite{peccianti}. In this context,
$d,g,q$ are $O(1)$ while $\nu$ is $O(10^2)$ \cite{peccianti,assanto_book}.

\begin{figure}[ht]
\centering
\includegraphics[scale=.33]{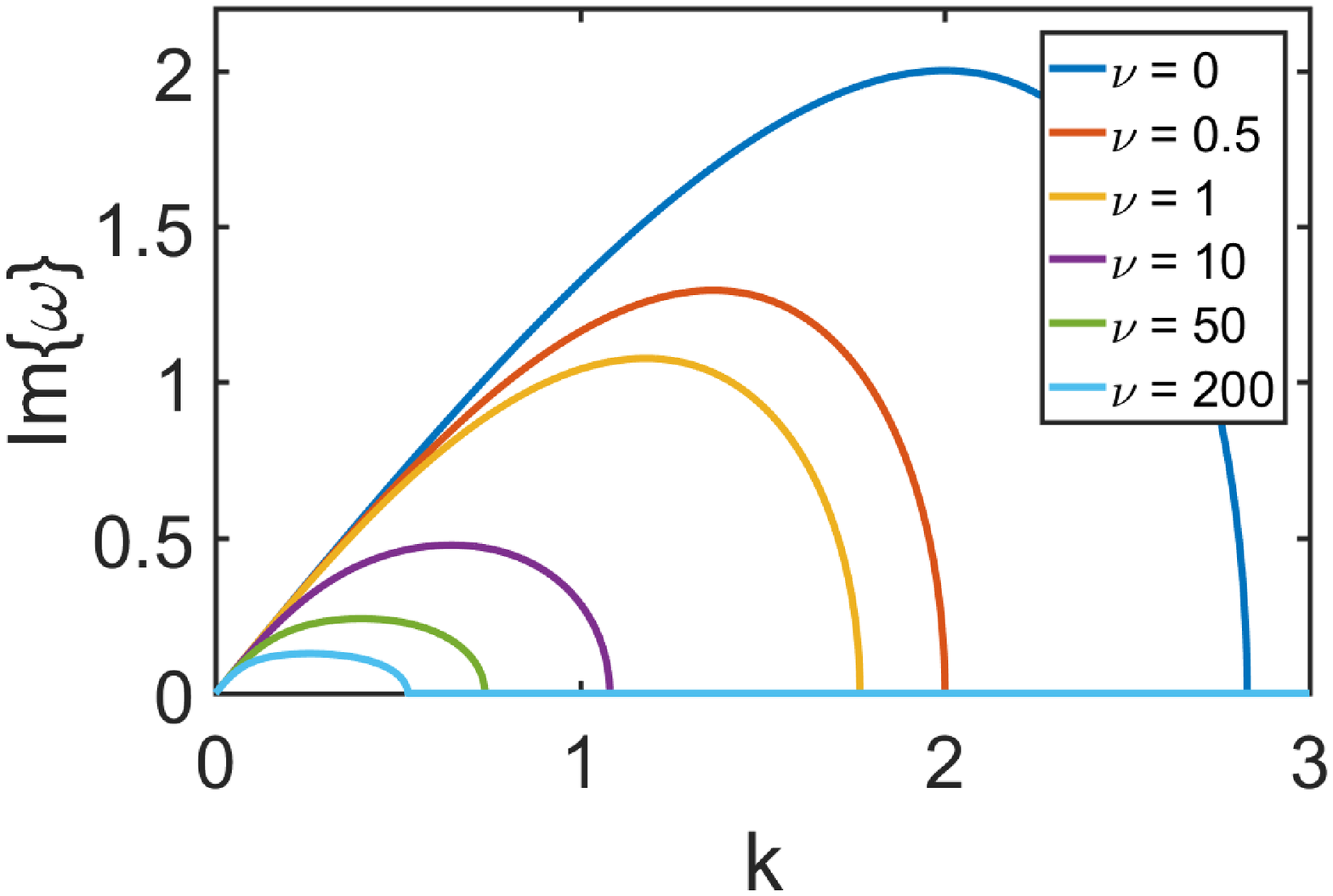}\\
\includegraphics[scale=.33]{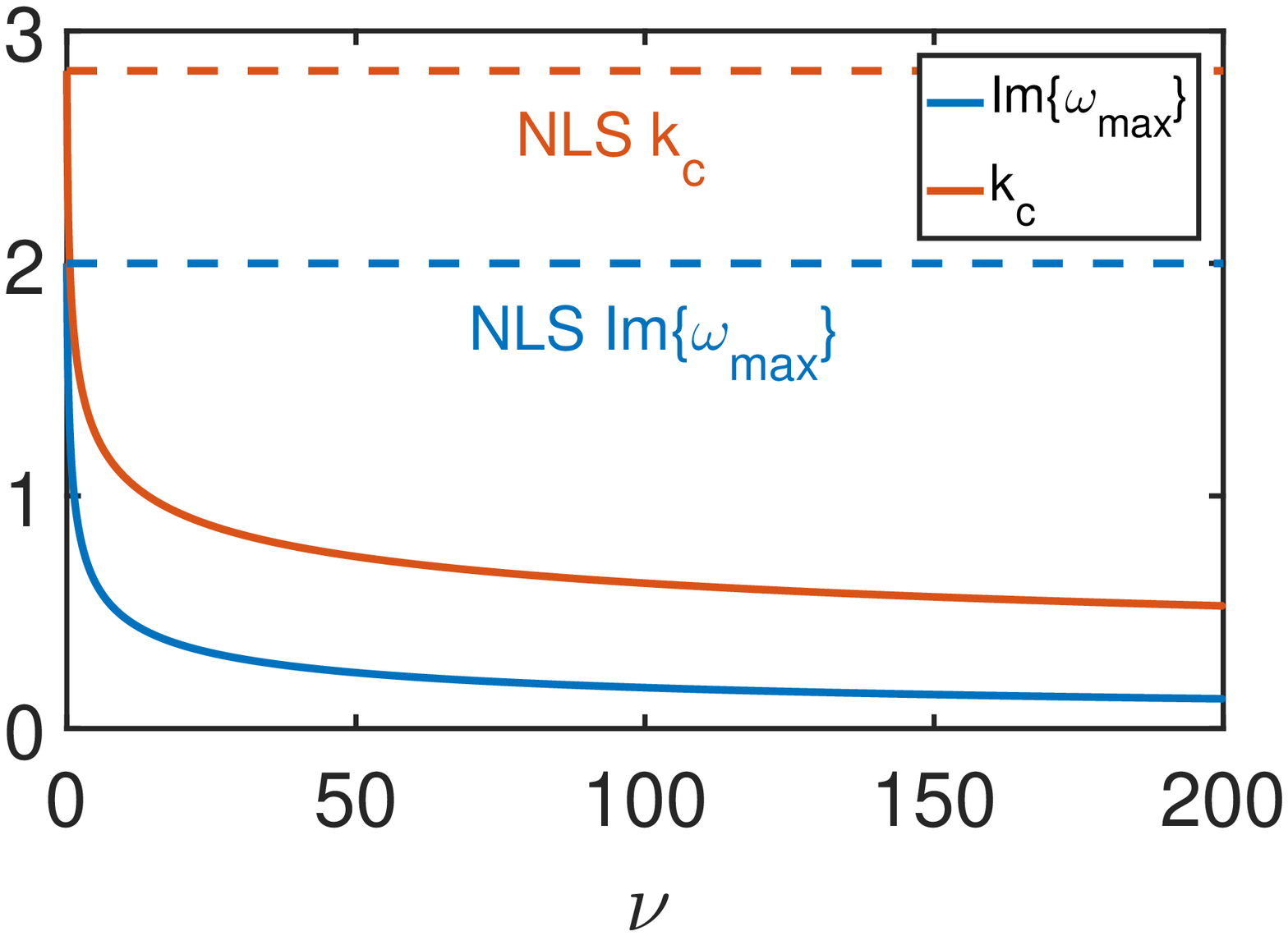}
\caption{(Color Online) Top: Growth rates for different values of the nonlocal
parameter $\nu$. Bottom: The change of critical values $\mathrm{Im}\{\omega_{\mathrm{max}}\}$
and $k_c$ with the nonlocality $\nu$.}
\label{mi_values}
\end{figure}

In order to investigate the stability properties of system \eqref{nls} consider its cw wave
solution
\[
u(z)=u_0e^{2ig\theta_0z},\theta_0=\frac{1}{q}u_0^2
\]
where $u_0$ is a real constant. Consider a small perturbation to this cw solution
\[
u(x,z)=[u_0+u_1(x,z)]e^{2ig\theta_0z}
\]
which is assumed to behave as $\exp[i(kx-\omega z)]$ provided the dispersion relation:
\begin{equation}
\omega ^2 = \frac{{d{k^2}\left( {d\nu {k^4} + 2dq{k^2} - 8gu_0^2} \right)}}{{\nu {k^2} +
2q}}
\label{dispersion}
\end{equation}
It is clear that when $dg>0$ the system is unstable (and is termed focusing) whereas when $dg<0$
the system is stable (and is termed defocusing). Also, when $\nu=0$ the equation reduces to the
dispersion relation of the relative NLS equation, which has the same stability criteria. From this
dispersion relation we can identify three critical values that characterize the instability, namely
the maximum growth rate, $\mathrm{Im}\{\omega_{\mathrm{max}}\}$ and its location,
$k_{\mathrm{max}}$, and the width of the instability region, $k_c$. The first,
$\mathrm{Im}\{\omega_{\mathrm{max}}\}$, is a measure of the propagation distance needed for the
instability to occur (the higher its value the faster the instability occurs) and the last, $k_c$,
defines the range of possible wavenumbers that can deem the system unstable (the higher its value
the more unstable the system as more cw's can cause an unstable propagation). By differentiating
Eq. \eqref{dispersion}, with respect to $k$, we find that $k_{\mathrm{max}}$ is the solution of the
algebraic equation
\[
d{\left( {\nu {k^3} + 2qk} \right)^2} - 8gqu_0^2 = 0
\]
while $k_c$ satisfies
\[
{d\nu {k^4} + 2dq{k^2} - 8gu_0^2}=0
\]
Both equations can be solved in closed form (they are bi-quadratics) to give the relative
dependance of $\mathrm{Im}\{\omega_{\mathrm{max}}\}$ and $k_c$ with the nonlocality $\nu$. We
illustrate this in Fig. \ref{mi_values}. Hereafter we fix $d=1/2$ and $g=q=u_0=1$.

This figure confirms the findings of Ref. \cite{kroli} that the nonlocality has a stabilizing
effect in the system. Indeed, both critical values that characterize the instability,
$\mathrm{Im}\{\omega_{\mathrm{max}}\}$ and $k_c$, decrease as $\nu$ increases. This means that the
effect will need more distance to be exhibited (and if $\nu$ is large enough this distance can be
larger than the experimental scales) and that a smaller range of wave numbers will cause an
instability. Notice, again, that while both values decrease,the effect, in the focusing case, is
always present, just suppressed. The limiting NLS system is, by these values, significantly more
unstable.

\begin{figure}[ht]
\centering
\includegraphics[scale=.2]{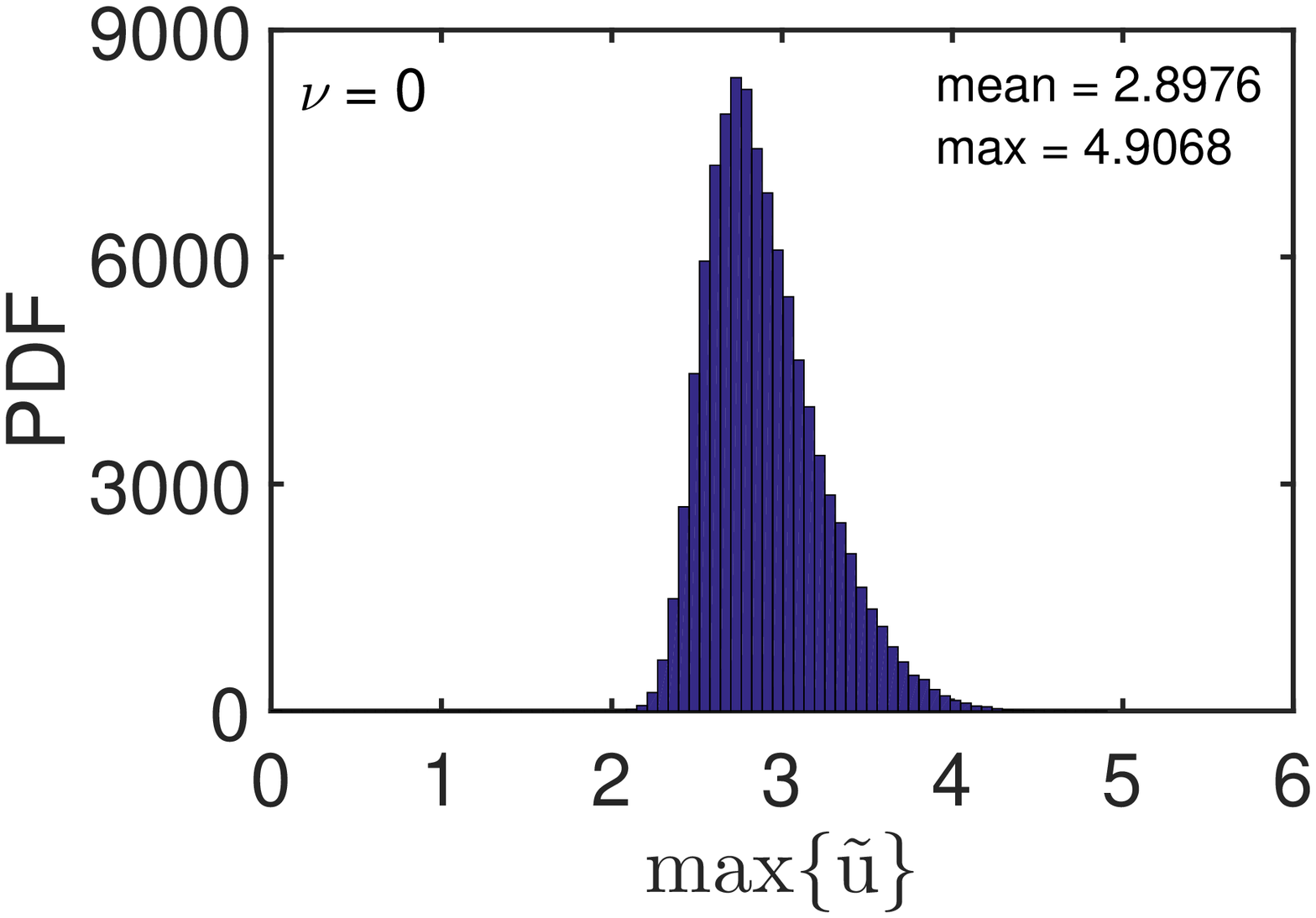}
\includegraphics[scale=.2]{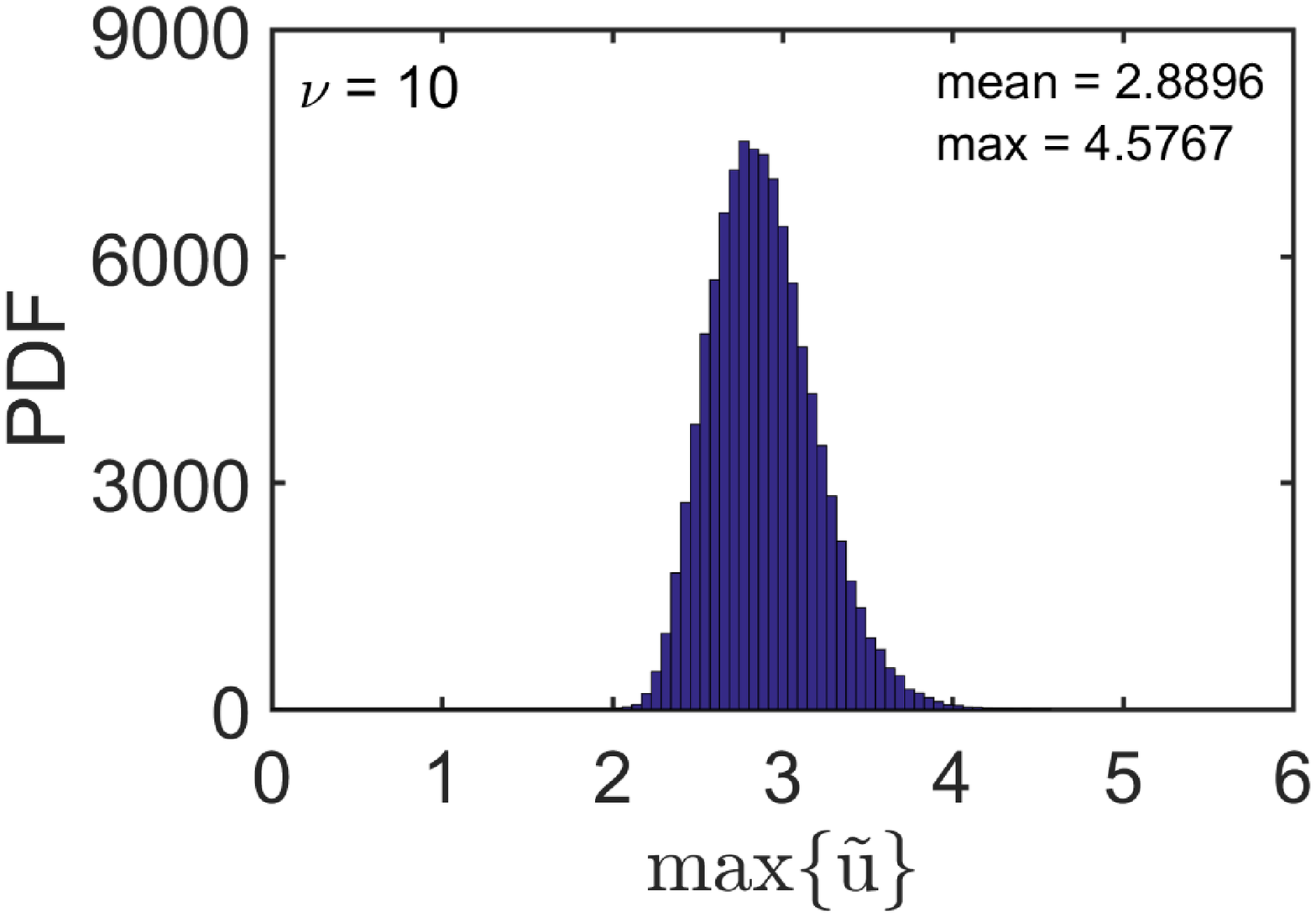}
\\
\includegraphics[scale=.2]{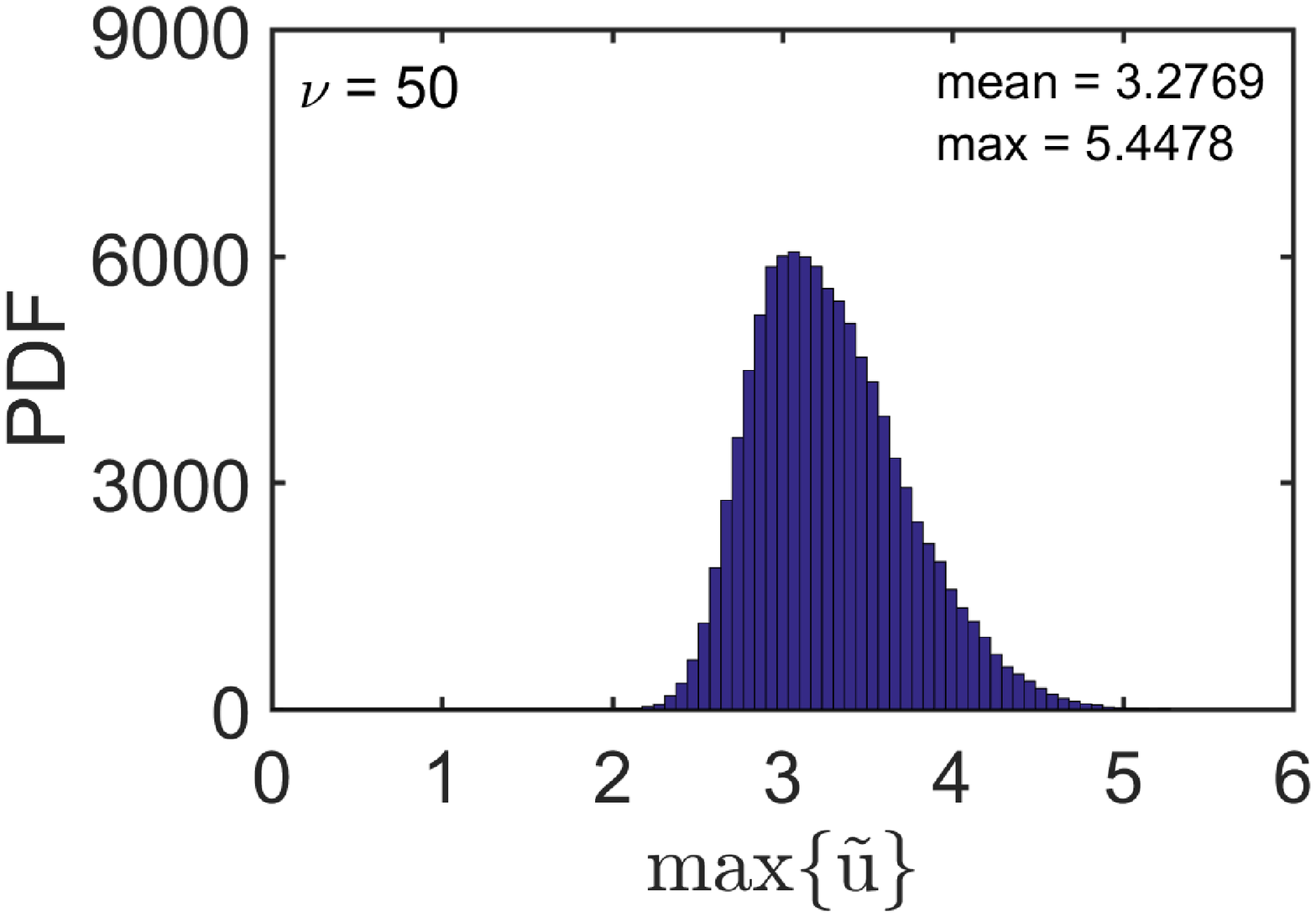}
\includegraphics[scale=.2]{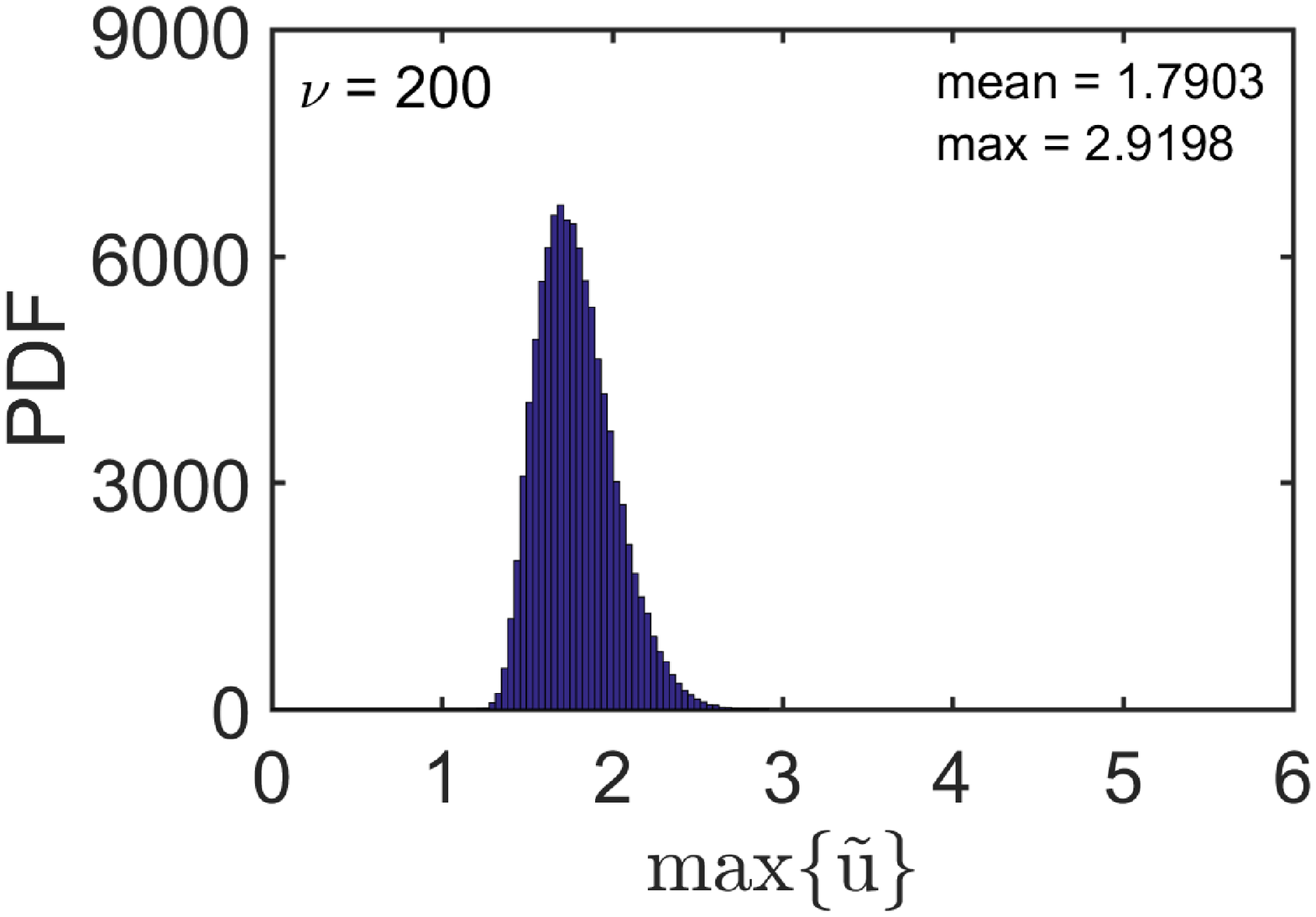}
\caption{(Color Online) Probability density functions of the maximum value
or $\max\{\tilde u\}$ for different values of the nonlocal parameter $\nu$.}
\label{pdf}
\end{figure}

To see how these observations affect the generation of rogue waves, we integrate numerically Eq.
\eqref{nls} using a pseudospectral method in space and exponential Runge-Kutta for the evolution
\cite{kassam} in a computational domain $x \in[-100,100]$, $z\in[0,20]$. An appropriate initial
condition would be a wide gaussian of the form
\[
u(x,0)= v(x,0)=e^{-x^2/2 \sigma^2},\; \sigma=30
\]
perturbed with additional 10\% random noise. A wide gaussian with randomness added is a prototype
of a set of broad/randomly generated states which can potentially excite more that one wave numbers
as it can be regarded as a Fourier series of different cw's of different $k$'s. This is
particularly important here as a single cw initial condition may not cause any growth due to the
decrease of $k_c$ with $\nu$. For each value of the parameter $\nu$ we perform $10^5$ trials. In
each trial we measure the highest wave amplitude and introduce the quantity
\[
\tilde u(x,z)  = \frac{u(x,z)}{\mathrm{max}\{ u(x,0) \}}
\]
which  measures the relative growth in amplitude from an initial state. Here we consider a rogue
event as one in which $\tilde u(x,z)$ at some value of $z$ is at least three times its maximum
initial value. In Fig. \ref{pdf} we depict the change in rogue events for some values of $\nu$.

These PDFs indicate that there is a relationship between the occurrence of rogue events and
nonlocality. Indeed, starting with $\nu=10$ the mean of the PDF is comparable to that of the
regular NLS ($\nu=0$). With $\nu=50$ there is a definite shift towards the right indicating that
rogue events have increased in both numbers and severity (amplitude). Finally, for $\nu=200$ there
is a sharp decrease of events and their amplitudes. This indicates that there is a nontrivial
dependence between the nonlocality and the occurrence of rogue events. The expectation that
nonlocality stabilizes the system and thus suppresses extreme phenomena does not hold.

\begin{figure}[ht]
\centering
\includegraphics[scale=.33]{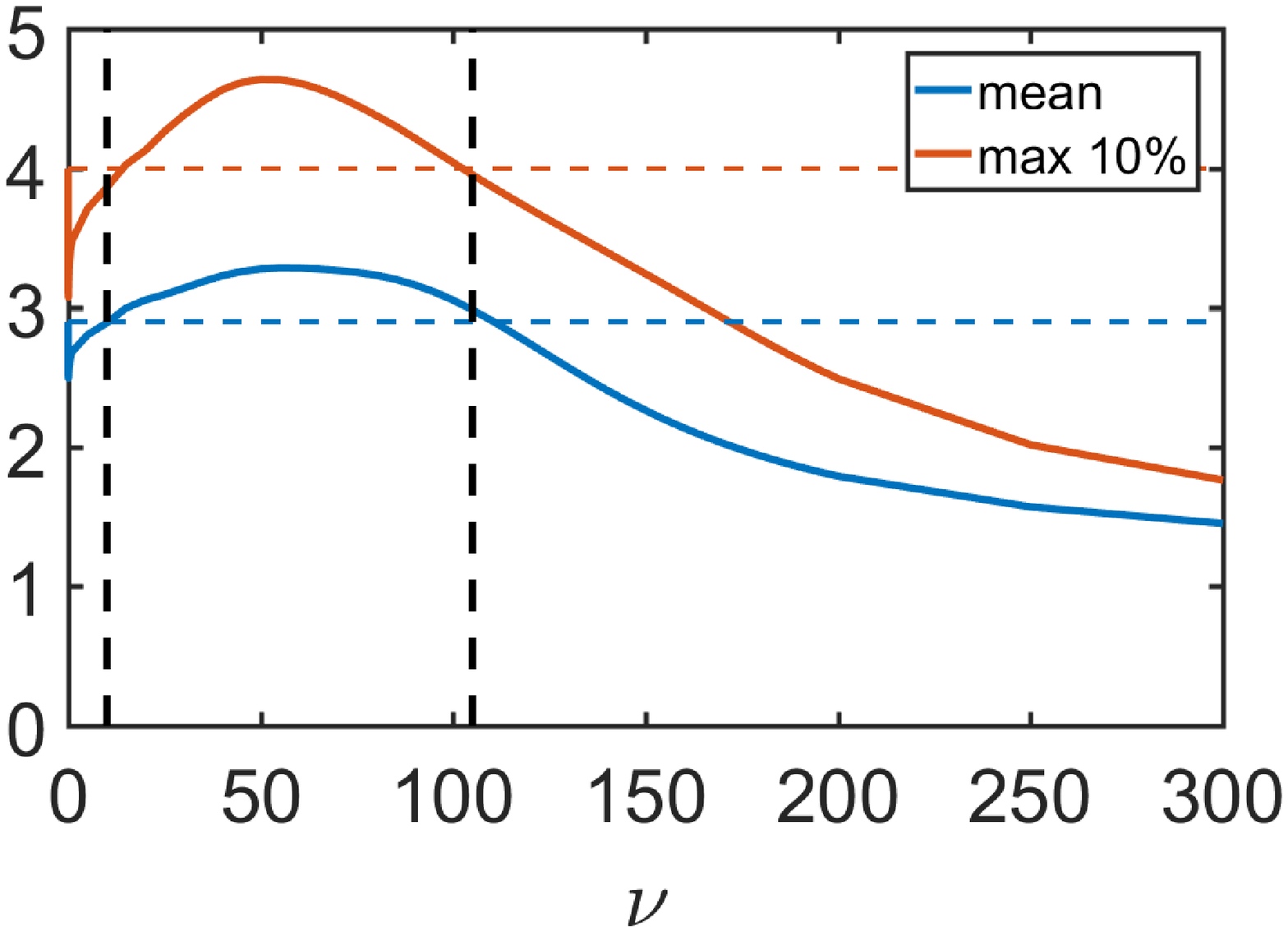}
\includegraphics[scale=.33]{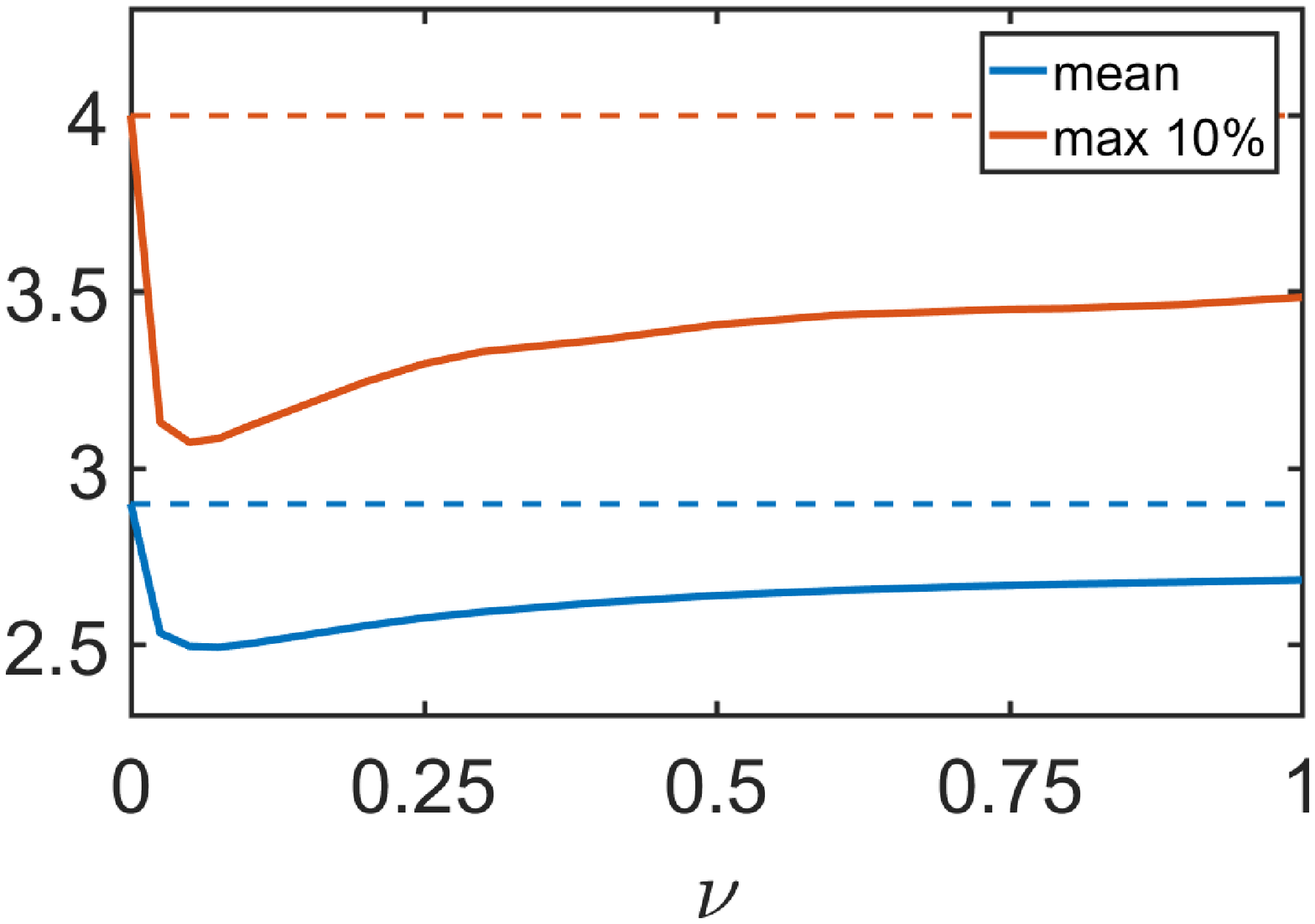}
\caption{(Color Online) Top:  The mean value of the PDFs and the mean value of the max 10\% events
with the nonlocal parameter $\nu$. The horizontal dashed lines indicate the relative values for
$\nu=0$ and the vertical dashed lines the values of $\nu$ for which these values surpass the NLS
system. Bottom: A zoom in around $\nu=0$.}
\label{mean_max}
\end{figure}

To further investigate the dependence of rogue events with $\nu$, we perform the same analysis for
a wide range of the parameter. In Fig. \ref{mean_max} we depict the change of the mean value in the
PDFs for the maximum values of $\tilde u$ with $\nu$ as well as the change of the top $10$\% of the
highest valued events.

Based on this figure, there are three different regions of interest. At first, when $0\leq
\nu\lesssim 10$, there is a sharp drop from the NLS case ($\nu=0$) to about $\nu=1$ and then both
curves increase with $\nu$, but still remain below the NLS limit. Of particular interest is the
transition from $\nu=0$. While we have taken points of order $10^{-2}$ in $\nu$ (in this region)
there is a sharp drop, a boundary layer type change, to a local minimum after which both the mean
and max curves increase. Next, in the region $10 \lesssim \nu \lesssim 110$, the curves remain well
above the NLS limits which translates into the system producing more numerous and more extreme
events. Recall, again that for these values the system exhibits very weak growth rates and has a
very narrow instability band. Finally, for $\nu>110$ the expected behavior is observed namely both
curves slowly decay as the nonlocal parameter increases.

\begin{figure}[ht]
\centering
\includegraphics[scale=.33]{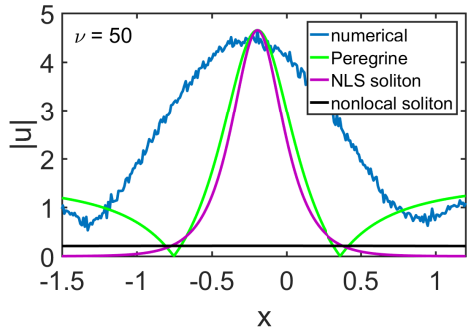}
\caption{(Color Online) Comparison of a (randomly chosen) rogue event of the nonlocal equation
with the known soliton and rational solutions.}
\label{compare}
\end{figure}

Next, considering the region where rogue wave are maximized, we now turn our attention to the
nature of these waves. Indeed, an important aspect of rogue wave formation is the type or shape of
the event, frequently  modeled by the so-called Peregrine soliton, a rational solution which reads
for Eqs. \eqref{nls} (with $\nu=0$)
\[
u_P(x,z)={u_0}
\left[ {1 - \frac{{4d{q^2} + i(16dgqu_0^2)z}}{{d{q^2} + (4gqu_0^2){x^2} + (16d{g^2}u_0^4){z^2}}}}
\right] {e^{2igu_0^2z/q}}
\]
while the single soliton solution is
\[
u_s(x,z)={u_0}{\operatorname{sech}}(u_0\sqrt{g/dq}x){e^{iu_0^2gz/q}}
\]
It is counter intuitive (and verified below) to believe that either would be a good candidate to
approximate rogue waves in this context as they lack the dependence on the nonlocal parameter
$\nu$. Furthermore, the soliton solution of Eqs. \eqref{nls} is \cite{mcneil}
\[
u(x,z) = \frac{{3q}}{2}\sqrt {\frac{d}{{g\nu }}} {\operatorname{sech}^2}(\sqrt {q/2\nu }
x){e^{2idq/\nu z}}
\]
which while it obviously depends on $\nu$, it has fixed amplitude (much like $\chi^{(2)}$ materials
\cite{karamzin,buryak}) which decays with $\nu$. As such, this solution is again not an appropriate
candidate to model extreme events (higher nonlocality results in smaller soliton amplitudes). In
fact, solutions with a free parameter for this system have been found but only in the defocusing
case and under a small amplitude approximation technique \cite{horikis2}. To illustrate we compare
all these solutions to an arbitrary rogue event in Fig. \ref{compare}.

\begin{figure}[ht]
\centering
\includegraphics[scale=.2]{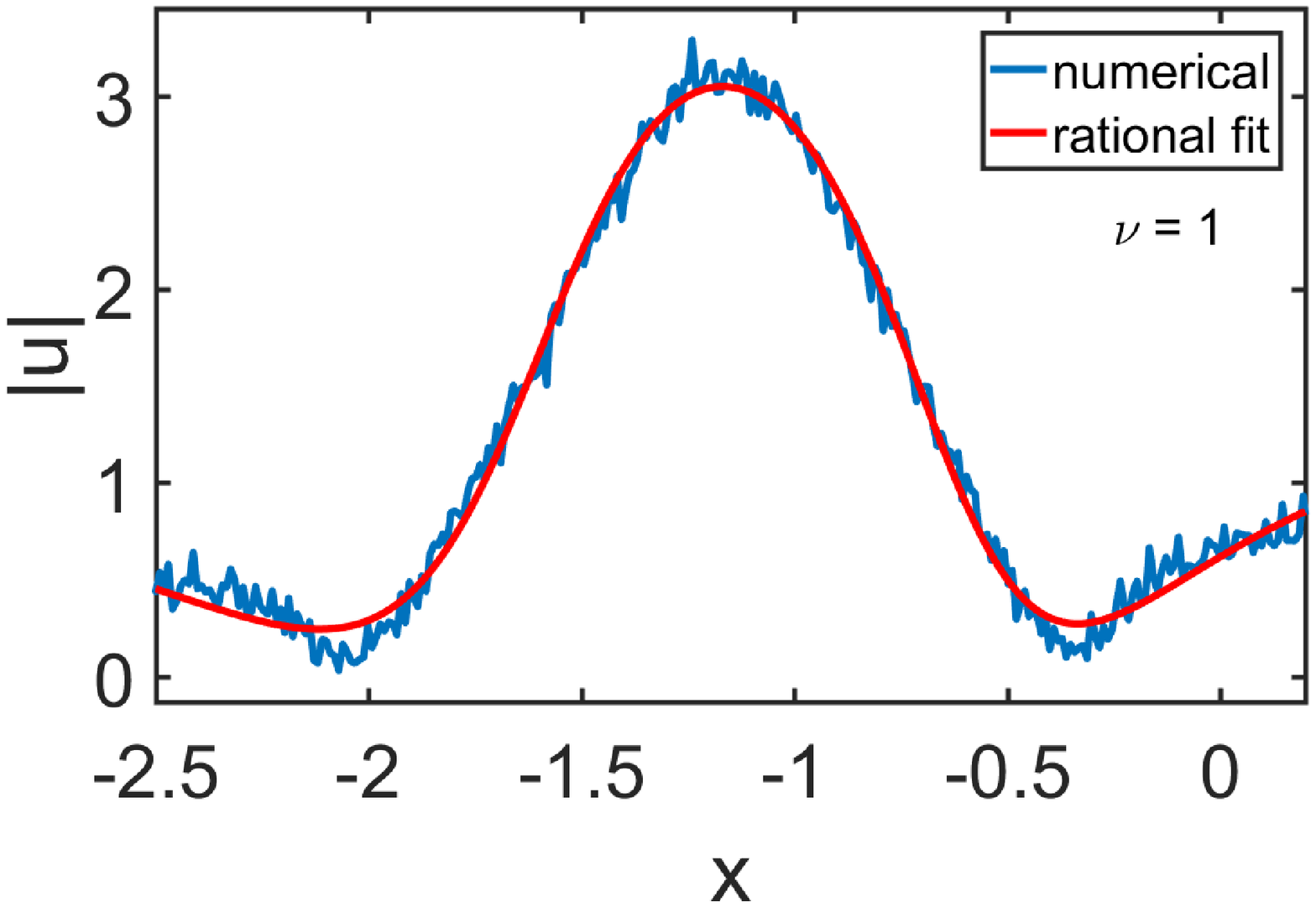}
\includegraphics[scale=.2]{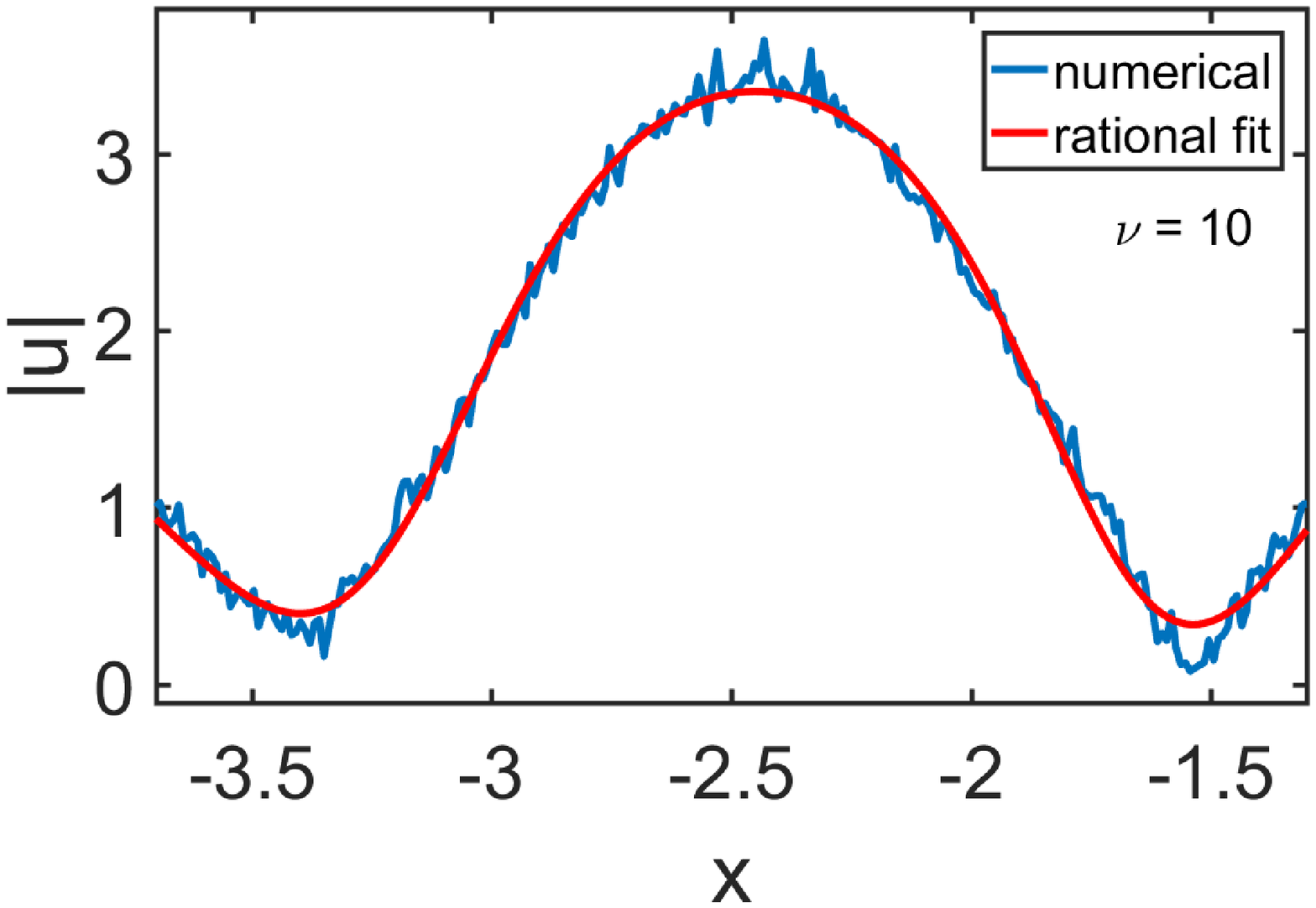}\\
\includegraphics[scale=.2]{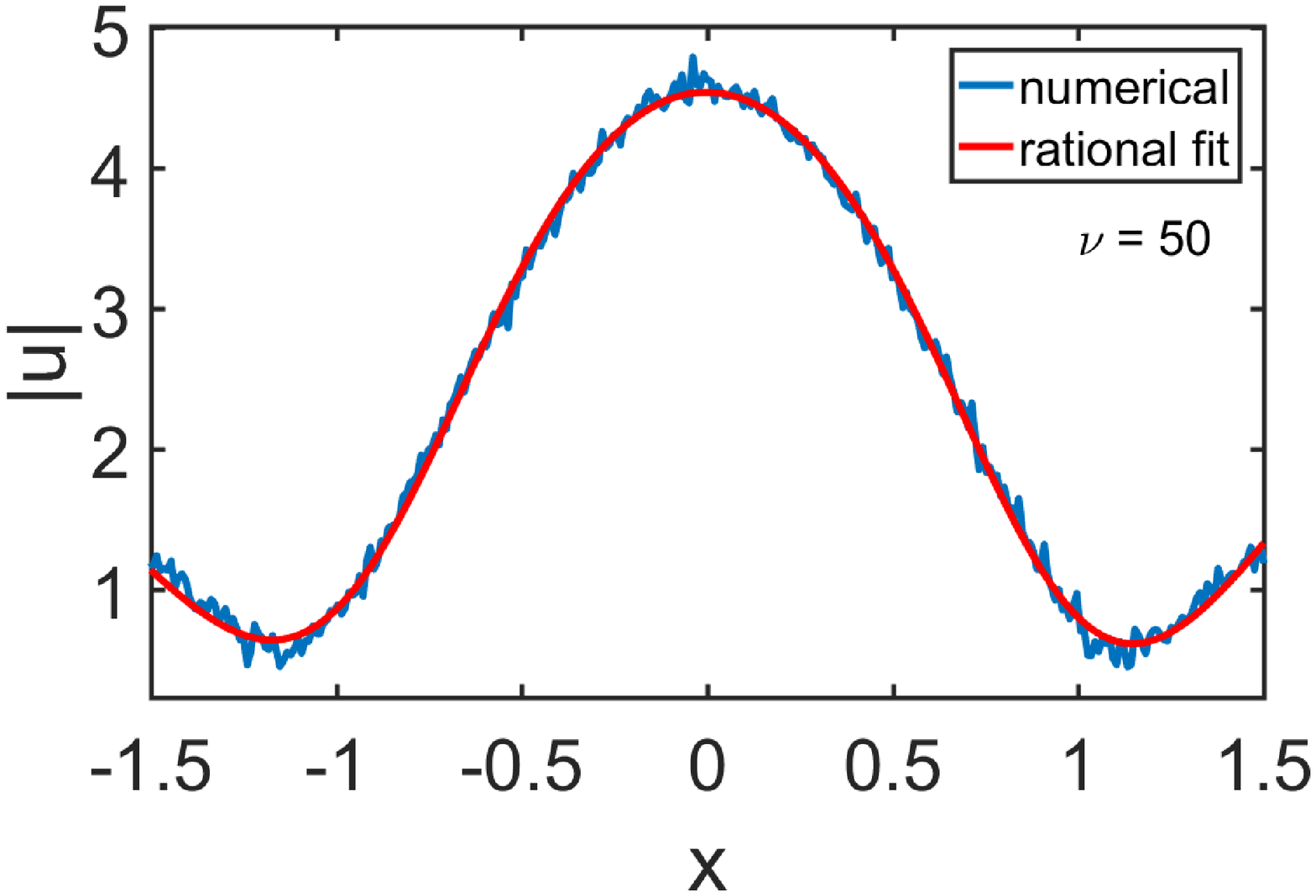}
\includegraphics[scale=.2]{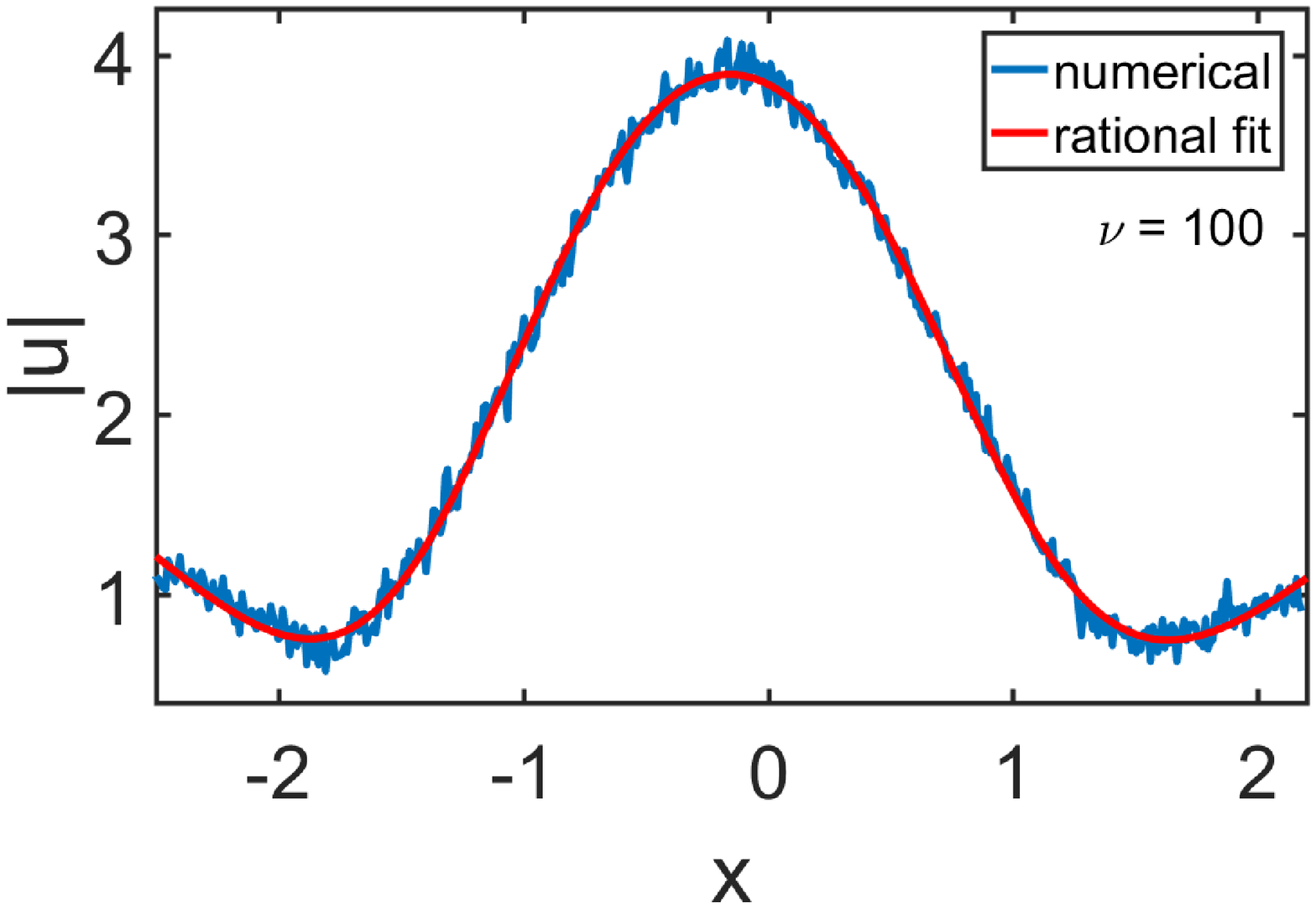}
\caption{(Color Online) A zoom in around a rogue event for the different values of the
nonlocal parameter $\nu$. A fourth order rational solution has been fitted (red line)
in all cases.}
\label{fits}
\end{figure}

Clearly the two solutions of the regular NLS system ($\nu=0$) are too narrow to fit the event,
while the decaying soliton of the nonlocal system is of the order 0.2 and appears as a straight
(black) line due to its magnitude. To further investigate the matter, in Fig. \ref{fits}, we zoom
in around a rogue event for different values of the parameter $\nu$ and fit a rational solution
around it.

The best fit is given by the ratio of two fourth order polynomials in $x$. We notice that the fits
become increasingly better as $\nu$ increases indicating the profound difference with the
integrable system. This is consistent with the soliton solutions. Indeed, the sech-type soliton of
the NLS is replaced by the sech$^2$-solution of the nonlocal system. This is not the first time
that more general (and commonly not known to be integrable) systems give rogue events whose nature
differs from that of the typical rational Peregrine soliton. A similar situation was recently
observed in deep water waves \cite{horikis}.

To conclude, we have studied rogue wave formation in nonlocal media using a physically important
nonlocal NLS system. For these systems, MI is suppressed in both the strength of growth rates and
size of instability band. Common belief suggests that this would also result in the appearance of
fewer and smaller, in amplitude, events. Contrary to that we found that for a wide range of values
of the nonlocal parameter, the system may produce significantly more events in both size and
numbers. The only known soliton solution of the system is not suited to describe these events which
also differ from their Kerr type counterparts in that they are well approximated by fourth order
rational solutions.

\begin{acknowledgments}
MJA is partially supported by NSF under Grant DMS-1310200 and the Air Force Office of Scientific
Research under Grant FA9550-16-1-0041.
\end{acknowledgments}

\end{document}